\documentclass[11pt, A4]{article}
\usepackage{graphicx}
\usepackage{epstopdf}
\usepackage{amsmath}
\usepackage{amssymb}
\usepackage{amsfonts}
\usepackage{amsthm}
\usepackage[usenames]{color}
\usepackage{array}
\usepackage{tabularx}
\usepackage{booktabs}           
\usepackage{caption}
\usepackage[usenames,dvipsnames,svgnames,table]{xcolor}
\usepackage[utf8]{inputenc}
\usepackage{lmodern}
\usepackage{verbatim}
\usepackage{amsmath}
\usepackage{amsfonts}
\usepackage{amssymb}
\usepackage{eso-pic}
\usepackage{graphicx}
\usepackage[final]{pdfpages}
\usepackage{pdfpages}
\usepackage[usenames,dvipsnames,svgnames,table]{xcolor}
\usepackage{mathtools}
\usepackage{empheq}

\usepackage[
      colorlinks=true,
      linkcolor=blue,
      urlcolor=blue,
      filecolor=blue,
      citecolor=red,
      pdfstartview=FitV,
      pdftitle={},
      pdfauthor={},
      pdfsubject={},
      pdfkeywords={},
      pdfpagemode=None,
      bookmarksopen=true
]{hyperref}

\usepackage{empheq}
\usepackage{mathtools}
\usepackage{epsfig}
\usepackage{cancel}
\usepackage{hyperref}

\def\beq{\begin{eqnarray}}
\def\eeq{\end{eqnarray}}

\def\be{\begin{equation}}
\def\ee{\end{equation}}
\def\bea{\begin{eqnarray}}
\def\eea{\end{eqnarray}}

\def\nn{\nonumber}

\definecolor{X}{rgb}{0,0,1}
\definecolor{Y}{rgb}{1,0,0}
\definecolor{Z}{rgb}{0,51,0}

\newcommand{\sbr}[1]{\left[#1\right]}

\newcommand{\br}[1]{\left(#1\right)}

\numberwithin{equation}{section}

\renewcommand{\thefootnote}{\fnsymbol{footnote}}

\begin{document}

\begin{centering}

\thispagestyle{empty}


{\LARGE \textsc{Milne Spacetime with conical defect: Some holographic Studies}} \\

 \vspace{0.8cm}

{\large 
Swayamsidha Mishra \footnote{We are still grieving the loss of our friend and collaborator Swayamsidha Mishra who passed away while this work was being finished. Her invaluable contributions and dedication to this work  will always be remembered.  }$^{1,3}$, Sudipta Mukherji$^{2,3}$
and 
Yogesh K.~Srivastava$^{1,3}$}
\vspace{0.5cm}

\begin{minipage}{.9\textwidth}\small  \begin{center}

${}^{1}$ National Institute of Science Education and Research (NISER), \\ Bhubaneswar, P.O. Jatni, Khurda, 
Odisha, India 752050 \\
  \vspace{0.5cm}
$^2$Institute of Physics, Sachivalaya Marg, \\ Bhubaneswar, Odisha, India 751005 \\
  \vspace{0.5cm}
$^3$Homi Bhabha National Institute, Training School Complex, \\ Anushakti Nagar, Mumbai India 400085 \\
  \vspace{0.5cm}
{\tt  mukherji@iopb.res.in, 
yogeshs@niser.ac.in}
\\ $ \ , $ \\

\end{center}
\end{minipage}

\end{centering}

\renewcommand{\thefootnote}{\arabic{footnote}}

\begin{abstract}

We initiate a holographic study of field theory in time dependent background with a conical defect.
We focus on the Milne spacetime 
to which, in the absence of cosmological constant, at late time  any hyperbolic Friedmann-Robertson-Walker 
metric flows to. When the Milne vacuum is represented by the adiabatic one,
we are able to compute the two point correlators of operators which are dual to the massive scalars in the
bulk AdS-Milne background with a defect. We find, for both twisted and untwisted operators, the
correlators can be represented as the sum over images. This sum can be
carried out explicitly to write the results in compact forms.


\end{abstract}

\newpage

\setcounter{equation}{0}
\section{Introduction}



\bigskip

\bigskip

AdS/CFT duality has been extremely useful in exploring various features of  strongly coupled quantum field theories 
as well as in providing insights into the quantum theory of gravity, 
particularly in the context of black hole physics. 
However, progress has been somewhat limited in exploiting the duality in 
the cosmological context. This could 
be due to the fact that a cosmological spacetime typically contains a space-like singularity and around
that region, classical gravity becomes unreliable. Nonetheless, 
since such a singularity generally runs all the way to the boundary, 
one may wish to examine if the boundary gauge theory could sense
this singularity. Indeed in \cite{Hertog:2005hu}-\cite{Chatterjee}, in the context of 
AdS cosmologies and in particular, for AdS-Kasner and 
AdS-FRW cosmologies, such questions were addressed with various degrees of success. 

Among the class of hyperbolic FRW geometries, 
the simplest is the Milne spacetime where the scale factor has a 
linear dependence on time. On one hand, since by a coordinate transformation 
the Milne spacetime can be expressed as the future wedge of the Minkowski metric, 
the geometry is free of curvature. On the other hand, as the 
time dependence of the metric is relatively simple,
one may hope to have some analytical handle \cite{Birrell:1982ix}, \cite{Saharian}. 
Further, in the absence of a cosmological constant, we note that all the hyperbolic FRW geometries
approach the Milne spacetime at late time as the expansions in these models drive the 
mass-energy density to zero.
In this paper, our focus will be on the Milne space-time.
Within the  holographic set up, we will represent the boundary by the Milne geometry. The corresponding 
bulk dual will be the AdS-Milne spacetime. The precise nature of the gauge theory on 
this boundary will depend on the spacetime dimensions. 

A generic difficulty that arises while working with quantum field theories 
on a curved, especially time dependent, 	background
is that the choice of the vacuum becomes ambiguous. Consequently, the correlation 
functions constructed out of the quantum fields start
to depend on the choice of the vacuum. Even though the Milne spacetime is a patch of Minkowski, 
there exists multiple complete sets of modes which are related by 
Bogoliubov transformations. Fock space built out of the corresponding 
creation and annihilation operators are necessarily inequivalent and 
corresponding vacuum states carry different physical properties. Two preferred choices 
for the vacuum states in the Milne spacetime are
the adiabatic
and the conformal vacuum. Other less commonly used vacua can also be
defined, see for example \cite{Akhmedov:2021agm}.  Among these vacua, the adiabatic vacuum 
is particularly appealing because of its similarity with the familiar vauum in the Minkowski spacetime.

In the present work, to start with, we perform several computations  in AdS-Milne background and 
extract the corresponding gauge theory quantities. For a massive minimally coupled scalar field
we compute the bulk Wightman and the other Green functions in the case when 
the Milne part of the vacuum is represented by the adiabatic one.
Subsequently, following the holographic prescription, the corresponding boundary 
correlators are extracted. An important issue here is whether 
the subregion duality \cite{Bousso:2012mh} holds. We find that, at least for the cases we study, 
the bulk and the boundary correlators
seem to respect the subregion duality. Next we turn our attention to the one where the 
Milne part is represented by the conformal vacuum. 
Here, we have not been able to find a closed form expression for the Wightman function. 
However, we see that the retarded Green function
has the same form as the one obtained in the adiabatic vacuum. This is indeed expected 
as  the retarded correlator 
is supposed to be a state independent function. Equipped with these computations, 
we turn our focus on the Milne spacetime  in the presence of a conical defect.

Field theory on spacetime with a conical defect has been of interest for long \cite{Sommerfeld} -
\cite{Solodukhin:1994yz}.
Lately, with the
advent of
the holographic correspondence, exploring the properties of a class of strongly coupled field theories on 
conical spacetime
has become possible \cite{Balasubramanian:2014sra} - \cite{Bayona:2010sd}.
For example, this correspondence can be exploited to compute the correlation functions of operators in a 
strongly coupled
field
theory, admitting large $N$ expansion, on a spacetime with a conical defect
once the gravitational dual is known. As a simple illustration,
let us consider a field theory on a three dimensional background given by
\begin{equation}
ds^2 = - dT^2 + dR^2 + R^2 d\theta^2,
\end{equation}
where the angular coordinate $\theta$ has a periodicity of $2\pi/q$. For any value of $q$ other than
$1$, there is a singularity at $R =0$. For $q > 1$, to which we would restrict to, the spacetime has an angle
deficit. In the large $N$ limit
of the field theory, the leading contribution to the two point correlation of a scalar operator would come
from examining appropriate scalar field on the AdS spacetime with the conical singularity.
Written in the Poincare coordinates, the bulk geometry in question is therefore
\begin{equation}
ds^2 = \frac{1}{z^2}\Big[dz^2 - dT^2 + dR^2 + R^2 d\theta^2\Big],
\label{adscone}
\end{equation}
where $z = \infty$ represents a horizon and $z = 0$ is the boundary where the gauge theory lives. The details
of this theory follow from the $M2$ brane of the M-theory compactified on $S_7$. Note that the singularity
present on the boundary now extends for all values of the radial coordinate $z$.
The Wightman function of a minimally coupled massive scalar  $\Phi$
of mass $m$ on this spacetime can be read out from \cite{deMello:2011ji}. 
Denoting the coordinates $(z, T, R, \theta)$
together as
$x$
and restricting to
the integer values of $q$,
it is given by
\begin{equation}
G_+(x,x') = \langle0|\Phi(x) \Phi(x')|0\rangle = -\frac{1}{4\pi^2} \sum_{k=0}^{q-1} \frac{Q_{\nu
-1/2}^1(w_k)}{\sqrt{w_k^2 -1}},\nn
\end{equation}
where $Q_{\nu -1/2}^1(w_k)$ is the associated Legendre function and
\begin{equation}
w_k = 1 + \frac{R^2 + R'^2 - 2 R R' ~{\rm cos}(\theta - \theta' - 2\pi k/q) + (z - z')^2 - (T - T')^2}{2zz'},\nn
\end{equation}
and $\nu = {\sqrt{9/4 + m^2}}$. On the boundary, this  field is dual to a scalar primary operator
$\cal{O}$ with scaling dimension $\Delta = \nu + 3/2$. The correlator can then be extracted following the
BDHM prescription \cite{Banks:1998dd}, namely,
\begin{eqnarray}
\langle \Psi| {\cal{O}}(T,R, \theta){\cal{O}}(T', R', \theta')|\Psi\rangle &=&  \lim\limits_{z,z'\rightarrow 0}
(zz')^{-(\nu +3/2)} \Big[-\frac{1}{4\pi^2} \sum_{k=0}^{q-1} \frac{Q_{\nu
-1/2}^1(w_k)}{\sqrt{u_k^2 -1}}\Big] \nonumber\\
&=& \sum_{k =0}^{q-1}\frac{C_\Delta}{[-(T-T')^2 + R^2 + R'^2 -2 R R' ~{\rm cos}(\theta - \theta' -
2\pi k/q)]^\Delta}\nonumber\\
\label{adsc}
\end{eqnarray}
where
\begin{equation}
C_\Delta = \frac{2^{2\Delta - 4} \Gamma(\Delta) \Gamma(\Delta -1)}{\pi^2 \Gamma(2\Delta -2)}.\nn
\end{equation}
Here $|\Psi\rangle$ is the appropriate boundary state of the field theory with the conformal symmetry
broken by the defect.
Though the sum above can be performed, written in this way, the correlator has an interpretation
in terms of the sum over images.

The metric (\ref{adscone}) can arise due to the presence of a cosmic string. In the weak field approximation and
in the thin string limit, the parameter $q$ gets related to the mass density of the string. If we wish to model the
formation of such a defect and examine the particle creation during the formation of the defect, 
we need to go beyond this static spacetime and replace it by an appropriate dynamical
one \cite{Davies, Sahni}. One possibility could be to consider a geometry of the form
\begin{equation}
ds^2 = \frac{1}{z^2}\Big[dz^2 - dT^2 + dR^2 + f(T) R^2 d\theta^2\Big],
\end{equation}
where $f(T) = [1 - ~{\rm tanh}(T/T_0)]/2$, and $\theta$ is a coordinate of period $2\pi$. It represents
a quench of the angular coordinate around $T = 0$, lasting for a time $T_0$. One may hope to gain some
insights into the behaviour of the strongly coupled field theory on the boundary by examining the bulk
representing
this dynamic cone. Difficulty arises  immediately however due to the absence of an analytical handle. For
example, for a generic
function $f(T)$, the Klein-Gordon equation would not admit a separation of variables of the field. In order
that to
happen we need to assume that the field is independent of the coordinate $\theta$. While, even with such a
cylindrical symmetry, it might be interesting to explore the dual field theory,
in this work we take a modest step. Here we intend to study 
aspects of strongly coupled gauge theory on the Milne geometry in the presence of a conical defect
\footnote{The possible occurrence of topological defects in the early universe and various cosmological
consequences due to their presence have been an active area of research in the past, see for example \cite{VS}.}   .  
With the choice of the adiabatic vacuum, the correlators of the operators dual to
the massive scalars in the bulk can be computed. The final form turns out to be similar to
the one given in  (\ref{adsc}) written in the Milne coordinates but we work this out explicitly
starting with the quantization of minimal massive scalar in AdS-Milne background. 
Because of the presence of the defect, on the boundary
we can have  twisted scalar operators that are
dual to the bulk scalar fields with twisted boundary conditions (\ref{tperiod}). It follows from \cite{CJI} and
subsequently from \cite{LHF} that the twisted scalars can be defined on a non-simply connected spacetime. 
These scalars satisfy the same equations of motion as that of the  
untwisted scalars but differ in their boundary conditions. We end our exploration with the computation of the boundary
correlators involving the twisted operators dual to these scalars.

\section{AdS-Milne}

We start out with the  computation of  the bulk and the boundary correlators in the AdS-Milne spacetime. 
This, in turn, will
set the stage for a similar, but more involved, calculation of correlators in AdS-Milne 
spacetime with a conical defect. This is analyzed in 
a subsequent section. 
Mode expasions of a massive scalar field in this geometry turn out to be sensitive to the spacetime dimensions. 
Therefore
we carry out our study both in four and five dimensions. The later has been provided in the appendix.

\subsection {$3+1$ dimensional AdS-Milne}

In the Poincare coordinates, the AdS-Milne metric in four dimensions takes the form 
\bea
ds^2 = \frac{1}{z^2}\left( -dt^2+t^2dr^2+t^2\sinh^2 r d\theta^2 +dz^2 \right),
\label{milnethree}
\eea
$\theta$ being a periodic coordinate with  period $2\pi$.
We see that we have a Milne spacetime for every value of the bulk radial coordinate $z$. 
The fact that the AdS-Milne is a subregion or 
a patch in the AdS can be seen from   
the coordinate transformations, $T = t\cosh r,~R = t\sinh r $. These transformations cover only the part
$T \ge 0, R \ge 0$ of the AdS. 
Even though the Poincare AdS and AdS-Milne written in Poincare coordinates are related by coordinate transformations, 
we must be careful while studying field theory
in these backgrounds  as corresponding propagators need not be related by similar coordinate transformations. 
Therefore, in the holographic set up, 
we should independently define the Green functions in each coordinate system.

The equation of motion of a minimally coupled scalar of mass $m$ on AdS-Milne
is
\begin{equation}
\frac{1}{\sqrt{-g}}\partial_\mu (\sqrt{-g} g^{\mu\nu} \partial_\nu \phi) - m^2 \phi = 0,
\end{equation}
Written explicitly, it takes the form
\begin{eqnarray}
&&- \frac{2 z^2}{t} \partial_t \phi - z^2 \partial_t^2 \phi + \frac{z^2}{t^2} {\rm coth}r \partial_r
\phi + \frac{z^2}{t^2} \partial_r^2\phi + \frac{z^2}{t^2 {\rm sinh}^2r} \partial_\theta^2 \phi - 2 z \partial_z 
\phi + z^2 \partial_z^2\phi - m^2 \phi = 0.\label{kgeqn}
\end{eqnarray}
For generic values of $\nu = \sqrt{m^2 + 9/4}$, the solutions are either
\begin{eqnarray}
\phi_{\lambda \alpha n} (z, t, r, \theta) = C_{\lambda \alpha n}\Big[z^{\frac{3}{2}} J_{\nu}(\lambda z) \Big]
\Big[ \frac{H_{i\alpha}^{(2)}(\lambda t)}{\sqrt t}\Big]
P^{-n}_{i\alpha-\frac{1}{2}}({\rm
cosh}r)e^{in\theta},
\label {solutionseq}
\end{eqnarray}
or
\begin{eqnarray}
\phi_{\lambda \alpha n} (z, t, r, \theta) = \tilde C_{\lambda \alpha n}\Big[z^{\frac{3}{2}} 
J_{\nu}(\lambda 
z) 
\Big]
\Big[ \frac{J_{-i\alpha}(\lambda t)}{\sqrt t}\Big]
P^{-n}_{i\alpha-\frac{1}{2}}({\rm
cosh}r)e^{in\theta}
\label{solutionseq2}
\end{eqnarray}
In these equations, $J_\mu(x), H_{i\alpha}^{(2)}(\lambda t)$ and $P_{i\alpha-1/2}^{-n} (u)$ are the Bessel 
function, Hankel 
function of the second kind and the associated Legendre function respectively. The constants $\lambda \ge 0$,
$\alpha \ge 0$ and $n$ takes integer values.
In writing down these solutions, we used the boundary conditions that the solutions are
regular at $z =0$ and $r = 0$. Since these solutions are normalizable modes, they are dual  to operators of conformal dimension $\Delta= \Delta_{+} = 3/2 + \nu $. Looking at the large $t$ behaviour of the Bessel and the Hankel functions, it is 
easily 
seen that that the choice $H_{i\alpha}^{(2)}$ is similar
to working with the Minkowskian positive energy modes. The vacuum  defined with respect to
these modes is called the adiabatic vacuum. On the other hand,  if we take $J_{-i\alpha}(\lambda t)$,
we are in the conformal vacuum \cite{Birrell:1982ix}. We first work with the former. The normalization constants 
$C_{\lambda 
\alpha n}$ are determined by the following condition, 
\begin{eqnarray}
\int dz dr d\theta {\sqrt{-g}}g^{tt} [\phi_{\lambda \alpha n} (z, t, r, \theta)
\partial_t\phi^*_{\lambda' \alpha' n'} (z, t, r, \theta)
&-& \phi^*_{\lambda' \alpha' n'} (z, t, r, \theta)
\partial_t\phi_{\lambda \alpha n} (z, t, r, \theta)] \nonumber\\
&&= -i \delta(\lambda - \lambda') \delta(\alpha - \alpha')
\delta_{nn'}. 
\end{eqnarray}{\label{normalization}
Using (\ref{solutionseq}), we get
\begin{equation}
C_{\lambda \alpha n} =  i\sqrt{ \frac{ {\alpha\lambda ~{\rm sinh}\pi\alpha}}{2\pi} } ~\Gamma[i\alpha + 1/2 +
n]\frac{e^{\frac{\pi\alpha}{2}}}{2}.
\end{equation}
To get to this, we have used the property (\ref{norma2}).
The canonical quantization proceeds by defining the field operator
\begin{equation}
\Phi = \sum_{i} [ \phi_i ~a_i + \phi_i^* ~a^\dagger_i],\nonumber
\end{equation}
where ${a_l}^\dagger, a_l$ are the creation and the  annihilation operators respectively and $i$ 
includes the set of quantum numbers $\lambda, \alpha, n$. The summation above represents integrations over
$\lambda, \alpha$ and a sum over $n$.

We start by computing the Wightman function. It is defined as
\begin{equation}
G_+(z,t,r,\theta ; z',t',r', \theta') = \langle 0|\Phi(z,t,r,\theta) \Phi( z',t',r', \theta')|0\rangle
= \sum_i \phi_i (z,t,r,\theta) \phi_i^*( z',t',r', \theta'). \label{wightman}
\end{equation}
Here, $|0\rangle$ includes the Minkowski vacuum for the Milne part. The other Wightman function
can be obtained from the relation $G_-(x,x') = G_+^*(x,x')$.
Denoting the set of coordinates $(z,t,r,\theta)$ together as $x$, we get
\begin{eqnarray}
G_+(x, x') &=&  \sum_{n=-\infty}^\infty\int_0^\infty d\lambda d\alpha 
\phi(z,r,t,\theta)\phi^*(z',r',t',\theta') = \frac{1}{4} \int d\lambda d\alpha
\Big(\frac{(z z')^{\frac{3}{2}} }{\sqrt{tt'}} \alpha\lambda ~{\rm sinh}(\pi\alpha) e^{\pi \alpha}\Big)\nonumber\\
&\times& J_\nu(\lambda z) J_\nu(\lambda z') H_{i\alpha}^{(2)}(\lambda t) H_{-i\alpha}^{(1)}(\lambda t')
\sum_{n=-\infty}^\infty \Gamma[i\alpha + 1/2 + n] \Gamma[-i\alpha + 1/2+n]\nonumber\\
&\times& P_{i\alpha - 1/2}^{-n} ({\rm{cosh}}r) P_{i\alpha - 1/2}^{-n} ({\rm{cosh}}r')
e^{in(\theta - \theta')}\nonumber\\
&=&\frac{(z z')^{\frac{3}{2}} }{\pi\sqrt{tt'}}\int_0^\infty d\lambda d\alpha
 \alpha\lambda ~{\rm tanh}(\pi\alpha)
J_\nu(\lambda z) J_\nu(\lambda z') K_{i\alpha}(i\lambda t) K_{i\alpha}(-i\lambda t')
P_{i\alpha - 1/2}({\rm cosh}\chi).\nonumber\\
\label{midexp}
\end{eqnarray}
The identities used to reach here have been presented in the appendix, see (\ref{appen1}).
In the above, $K_\nu(z)$ is the modified Bessel function and $\chi$ is defined as
\begin{eqnarray}
{\rm cosh}\chi = ~{\rm cosh}r~{\rm cosh}r' - ~{\rm sinh}r~{\rm sinh}r' ~{\rm cos}(\theta - \theta').\label{chi3d}
\end{eqnarray}

Equation (\ref{midexp}) can be further simplified using (\ref{appen2})
to arraive at
\begin{eqnarray}
G_+(x,x') &=& \frac{1}{\pi{\sqrt{2\pi}}} (zz')^{\frac{3}{2}} \int_0^\infty d\lambda 
\lambda^{\frac{3}{2}}
J_\nu(\lambda z) J_\nu (\lambda z')
\frac{K_{\frac{1}{2}} (\lambda {\sqrt{-t^2 - t'^2 + 2 t t' ~{\rm cosh}\chi}})}
{  {\sqrt{-t^2 - t'^2 + 2 t t' ~{\rm cosh}\chi}} }\nonumber\\
&=& -\frac{1}{4\pi^2 (u^2 -1)^{\frac{1}{2}}} Q_{\nu - 1/2}^1(u),
\label{bulkwightman}
\end{eqnarray}
where
\begin{equation}   
u = \frac{1}{2 z z'} \Big(-t^2 - t'^2 + 2 t t' ~{\rm cosh}\chi + z^2 + z'^2 + i\epsilon ~{\rm sgn}(t-t')\Big),
\label{geomilne}
\end{equation}
the geodesic distance between two points in AdS-Milne spacetime and $Q_{\nu - 1/2}^1(u)$ is the 
associated Legendre function. To arrive at the last line
of (\ref{bulkwightman}), we have used a result from \cite{bateman}. In the conformal limit, when the
mass $m^2 = -2$ (and consequently $\nu = 1/2$), the correlator simplifies. Owing to
the property that $Q_0^1 = -1/\sqrt{u^2 -1}$, we get 
\footnote{We have calculated the Wightman function for normalizable modes that is 
$G_{\Delta_{+}}$. There is a corresponding Wightman function $G_{\Delta_{-}}$ 
for the non-normalizable modes As shown in \cite{Dorn}, it's the sum $G_{\Delta_+} + G_{\Delta_-}$ 
which corresponds  to a simple power solution in the conformal limit . }

\begin{equation}
G_+(x,x^\prime) =
\frac{1}{4\pi^2(u^2-1)}.\nonumber
\end{equation}

\bigskip

\noindent{\underbar {Boundary correlator}}

\bigskip

Having obtained the bulk Wightman function, we can use the BDHM prescription to 
construct the boundary correlator. 
\begin{equation}
 \lim\limits_{z,z' \rightarrow 0} (zz')^{-(\nu+3/2)} 
\langle \Phi(z,r,t,\theta)\Phi(z', r',t', \theta') \rangle_{\rm AdS-Milne}
= \langle \Psi| {\cal{O}}(r,t,\theta) {\cal{O}}(r',t',\theta')|\Psi\rangle.\nonumber
\end{equation}
Here ${\cal{O}}$ is the scalar primary operator of dimension $\Delta = 3/2 + \nu$, dual to the bulk scalar $\Phi$. 
On the left the subscript indicates that the bulk computation is done on AdS-Milne, and on the right,
$|\Psi\rangle$ is the corresponding state of the boundary conformal theory. Implementing the above,
we get
\begin{eqnarray}
\langle \Psi|{\cal{O}}(t, r, \theta) {\cal{O}}(t', r', \theta')|\Psi\rangle
&=&  \lim\limits_{z,z' \rightarrow 0} (zz')^{-(\nu+3/2)}G_+(x,x')\nonumber\\
&=& \frac{(\Delta -2)2^\Delta}{4\pi^2 (-t^2 - t'^2 + 2 tt'~{\rm cosh}\chi)^\Delta}
\int_0^\infty \frac{{\rm cosh}\beta} {(1 + {\rm cosh}\beta)^{\Delta -1}} d\beta\nonumber\\
&=& \frac{C_\Delta}{(-t^2 - t'^2 + 2 tt'~{\rm cosh}\chi)^\Delta},
\label{bcorrelator}
\end{eqnarray}
where 
\begin{equation}
C_\Delta = \frac{2^{2\Delta -4} \Gamma(\Delta) \Gamma(\Delta -1)}{\pi^2 \Gamma(2\Delta -2)}.
\label{cdelta}
\end{equation}
To arrive at the final line, we have used (\ref{appen3}).

A similar computation can be carried out for the AdS-Milne spacetime in five dimensions
where the metric is 
\begin{equation}
ds^2 = \frac{1}{z^2}\sbr{-dt^2+t^2dr^2 +t^2\sinh^2 r d\theta^2 +t^2\sinh^2 r\sin^2\theta d\phi^2 +dz^2}.
\label{milnefour}
\end{equation}
The details of the computation is provided in the appendix. The Wightman function turns out to be
\bea
G_+(x,x^\prime) =  \frac{i}{(2\pi)^{5/2}(u^2-1)^{3/4}}   Q_{\nu-1/2}^{3/2}(u), \label{W.M.F-5}
\eea
where
\bea
u=\frac{1}{2zz^\prime}  (-t^2 - t^{\prime 2} + 2 tt^\prime~{\rm cosh}\gamma+z^2+z^{\prime 2}),\nonumber
\eea
leading to the boundary correlator
\bea
\langle\Psi| \mathcal{O}(t,r,\theta,\phi),\mathcal{O}(t^\prime, r^\prime, \theta^\prime,\phi^\prime)|\Psi\rangle
&=& \frac{2^{2\nu-1}\Gamma(\frac{1}{2}+\nu)\Gamma(\nu+2)}{(\pi)^{\frac{5}{2}}\Gamma(2\nu+1)}  
\frac{1}{a^{\nu+2}} \nn\\
&=&  \frac{2^{2\nu-1}\Gamma(\frac{1}{2}+\nu)\Gamma(\nu+2)}{(\pi)^{\frac{5}{2}}\Gamma(2\nu+1)}
\frac{1}{(-t^2-t^{\prime 2}+2tt^\prime \cosh \gamma)^{\nu+2}}\label{Milne-4B-Correlator}.\nn\\
\eea
Before proceeding to the next subsection, we end with the following note.
In a new coordinate system $\tau, \rho$,
defined as $t = b e^{\tau/b}$ and $r = \rho/b$ with $b$ being constant, the metric in (\ref{milnethree})
becomes
\begin{eqnarray}
ds^2 = \frac{1}{z^2} \Big[ dz^2 + e^{2\tau/b}\Big(-d\tau^2 + d\rho^2 + b^2 ~{\rm sinh}^2(\rho/b) 
d\theta^2\Big)\Big].\nonumber
\end{eqnarray}
Now, in the limit $b \rightarrow \infty$, the boundary becomes flat. The quantity $u$ defined in 
(\ref{geomilne})
reduces to $ u = \frac{1}{2zz'}[-(\tau -\tau')^2 + \rho^2 + \rho'^2 - 2 \rho\rho' ~{\rm cos}(\theta - \theta')]$
and the boundary correlator becomes the one in the Minkowski spacetime. 

\subsection{Comparison with Poincare-AdS}

Since the metric as well as the vacuum state are invariant under full Poincare-AdS symmetries, 
the final expression can be written in terms of the geodesic distance, 
which facilitates a comparison with the Green functions in Poincare-AdS. 

For AdS-Milne in 3+1 dimensions we get the geodesic distance $d(z,w)= \int_z^w ds$ as 
\bea
d(z_1,t_1,r_1,\theta_1;z_2, t_2,r_2,\theta_2) = \ln\frac{1+\sqrt{1-\xi^2}}{\xi},\nn
\eea 
where  $\xi$ is given by
\bea
\xi = \frac{2z_1z_2}{z_1^2+z_2^2 -t_1^2 -t_2^2  +2t_1t_2\cosh\chi}  =\frac{1}{u}.\nn
\eea
Using the relation between the hypergeometric function and associated Legendre function of second kind, namely 
\bea
Q_\nu^\mu(z) &=& \frac{e^{\mu\pi i}\Gamma(\nu+\mu+1)\Gamma(1/2)}{2^{\nu+1}\Gamma(\nu+3/2)}  
(z^2-1)^{\mu/2} z^{-\nu-\mu-1} ~ F\br{\frac{\nu+\mu+1}{2}, \frac{\nu+\mu+2}{2} ; \nu+3/2; \frac{1}{z^2}},\nn
\eea
we find,
\bea
Q_{\nu-1/2}^1\br{\frac{1}{\xi}} &=&  -4\pi^{2}  \br{\frac{1}{\xi^2}-1}^{1/2}G_{\nu + \frac{3}{2}}(\xi),\nn 
\eea
where 
\bea
G_{\nu+ \frac{3}{2}}(\xi) = \frac{2^{-\nu-5/2}\Gamma\br{\nu+3/2}}{  \pi^{3/2}\Gamma(\nu+1)}\xi^{\nu+3/2}   
F\br{ \frac{3}{4}+\frac{\nu}{2}, \frac{5}{4}+\frac{\nu}{2} ;  \nu+1 ; \xi^2 }\nn
\eea
is the scalar propagator for the normalizable modes in Poincare-AdS as given in \cite{Hoker}.

For AdS-Milne in 4+1 dimensions we can write the geodesic distance in a similar way 
\bea
d(z_1,t_1,r_1,\theta_1,\phi_1;z_2, t_2,r_2,\theta_2,\phi_2) = \ln\frac{1+\sqrt{1-\xi^2}}{\xi},\nn
\eea 
where $\xi$ is now given by
\bea
\xi = \frac{2z_1z_2}{z_1^2+z_2^2 -t_1^2 -t_2^2  +2t_1t_2\cosh\chi}  =\frac{1}{u}.\nn
\eea

Again, we can convert from associated Legendre functions to hypergeometric function
\bea
Q_{\nu-1/2}^{3/2}\br{\frac{1}{\xi}} =  -i(2\pi)^{5/2}  \br{\frac{1}{\xi^2}-1}^{3/4}  G_{\nu+2}(\xi),\nn
\eea
where 
\bea
G_{\nu+2}(\xi)  &=&  \frac{2^{-\nu-3} \Gamma\br{\nu+2}}{\pi^{2}\Gamma(\nu+1)}\xi^{\nu+2} 
F\br{ 1+\frac{\nu}{2}, \frac{3}{2}+\frac{\nu}{2} ;  \nu +1; \xi^2 }\nn
\eea
is the scalar Green function given in \cite{Hoker}.

\subsection{AdS-Milne in conformal vacuum}

As we discussed previously, working in conformal vacuum is equivalent to choosing $J_{-i\alpha}(\lambda t)$ 
in the $t$ part instead of the 
Hankel function. The basis of the mode expansion is therefore given by  
(\ref{solutionseq2}) rather than (\ref{solutionseq})
\bea
\phi(z,t,r,\theta)  = \tilde C_{\lambda \alpha n}\sbr{z^{3/2} J_\nu(\lambda z)} 
\sbr{\frac{J_{-i\alpha}(\lambda t)}{\sqrt{t}}}
P_{i\alpha-1/2}^{-n}(\cosh r)  e^{in\theta},
\label{phiconformal}
\eea
where the normalization constant, upto a constant phase factor, is given by
\bea
\tilde C_{\lambda \alpha n} =i\sqrt{\frac{\alpha\lambda}{4\pi} }~\Gamma\sbr{i\alpha +1/2 +n}.\nonumber
\eea

Now doing manipulations similar to what was done in the Minkowski vacuum, we arrive at
\bea
G_+(x,x^\prime) &=& \frac{1}{4\pi}\int d\lambda d\alpha \sbr{ \frac{(zz^\prime)^{3/2}}{\sqrt{tt^\prime}}
\alpha\lambda  }
J_\nu(\lambda z) J_\nu(\lambda z^\prime) J_{-i\alpha}(\lambda t)J_{i\alpha}(\lambda t^\prime)\nn\\
&&  \sum_{n=-\infty}^{+\infty}  \Gamma(i\alpha +1/2+n)\Gamma(-i\alpha +1/2+n)  P_{i\alpha-1/2}^{-n}(\cosh r)
P_{i\alpha-1/2}^{-n}(\cosh r^\prime) e^{in(\theta-\theta^\prime)}\nn\\
&=&  \frac{(zz^\prime)^{3/2}} {4\sqrt{tt^\prime}}\int d\lambda d\alpha \ \alpha\lambda
J_\nu(\lambda z) J_\nu(\lambda z^\prime) J_{-i\alpha}(\lambda t)J_{i\alpha}(\lambda t^\prime)\frac{1}{\cosh\pi\alpha}
P_{ i\alpha-1/2}(\cosh \chi)
\label{conformalone}
\eea
where $\chi$ is defined by \ref{chi3d}. The other Wightman function is 
given by $G_{-}(x,x^\prime) = G^{*}_{+}(x,x^\prime)$.

In $4+1$ dimensions, the right solution of the scalar equation of motion turns out to be
\bea
\phi(z,t,r,\theta,\phi)  = \tilde C_{\lambda\alpha}\sbr{z^{2} J_\nu(\lambda z)} 
\sbr{\frac{J_{-i\alpha}(\lambda t)}{t}}  Y_{\alpha lm}(r, \theta,\phi)\nonumber
\eea
where the normalization constant is given by 
\bea
\tilde C_{\lambda\alpha} =\sqrt{\frac{\pi\lambda}{2\sinh\pi\alpha} }.\nn
\eea
The Wightman function is then
\bea
G_{+}(x,x^\prime) =   \sum_{l,m} \int_0^\infty  \lambda. d\lambda d\alpha   \frac{\pi}{2}  
\frac{(zz^\prime)^2}{(tt^\prime)^2
\sinh\pi\alpha} J_\nu(\lambda z)J_\nu(\lambda z^\prime)
J_{-i\alpha}(\lambda t) J_{i\alpha}(\lambda t^\prime)   Y_{\alpha l m}(r,\theta,\phi)
Y_{\alpha lm}(r^\prime,\theta^\prime,\phi^\prime).\nn
\eea
As was done for the Minkowski vacuum, we can use the completeness relation of 
the spherical harmonics (\ref{completeness}) to cast
the above function as
\bea
G_{+}(x,x^\prime) =    \int_0^\infty  \lambda d\lambda d\alpha   \frac{1}{4\pi}  \frac{(zz^\prime)^2}{(tt^\prime)^2
\sinh\pi\alpha} J_\nu(\lambda z)J_\nu(\lambda z^\prime)
J_{-i\alpha}(\lambda t) J_{i\alpha}(\lambda t^\prime)\frac{\alpha \sin \alpha\gamma}{2\pi^2\sinh\gamma}. \nn
\label{conformaltwo}
\eea
where $\gamma$ is given, as before, by (\ref{xidef}).
Unfortunately, we have not succeeded in carrying out all the integrals in 
(\ref{conformalone}) and (\ref{conformaltwo}). 
However, we could make some progress while considering the retarded propagator. This is
given by $G_{R}(x,x') = \theta (t-t')G(x,x')$ where $G(x,x')= G_{+} (x,x') - G_{-}(x,x')$, 
difference between the  positive and the negative frequency Wightman functions.

To explicitly calculate the retarded propagator, we first consider the $3+1$ dimensional
AdS-Milne spacetime in conformal vacuum. In this case, we are able to write
\bea
G(x,x') = \int d\lambda d\alpha \frac{(zz^\prime)^{3/2}}{4\sqrt{tt^\prime}}  J_\nu(\lambda z)
J_\nu(\lambda z^\prime) \Big(J_{-i\alpha}(\lambda t)J_{i\alpha}(\lambda t^\prime)
-J_{-i\alpha}(\lambda t^\prime)J_{i\alpha}(\lambda t)\Big) 
\frac{\alpha\lambda P_{ i\alpha-1/2}(\cosh \chi) }{\cosh \pi\alpha}.\nonumber
\eea
We can straightforwardly show that the $G(x,x')$ in this vacuum is same as that in the Minkowski vacuum.
To this end,  we first write the Bessel function in terms of the Hankel function using $J_{\alpha}(x) =
\frac{1}{2}\left( H^{(1)}_{\alpha} (x) +  H^{(2)}_{\alpha} (x) \right)$.
Further, since  $H^{(1,2)}_{-\alpha} = e^{\pm i\pi \alpha} H^{(1,2)}_{\alpha}$, we can then write
\bea
4 \sbr {J_{-i\alpha}(\lambda t)J_{i\alpha}(\lambda t^\prime)- 
J_{-i\alpha}(\lambda t^\prime)J_{i\alpha}(\lambda t)} =
\sinh\pi \alpha \left( H^{(2)}_{i\alpha}(\lambda t)
H^{(1)}_{i\alpha}(\lambda t')-H^{(2)}_{i\alpha}(\lambda t')H^{(1)}_{i\alpha}(\lambda t) \right). \label{bessel}
\eea
Therefore,
\bea
G(x,x') &=& \int d\lambda d\alpha \frac{(zz^\prime)^{3/2}}{\sqrt{tt^\prime}}\alpha\lambda
J_\nu(\lambda z) J_\nu(\lambda z^\prime)  \times \nonumber \\
&&\tanh\pi\alpha P_{ i\alpha-1/2}(\cosh \chi)  \left( H^{(2)}_{i\alpha}(\lambda t)
H^{(1)}_{i\alpha}(\lambda t')-H^{(2)}_{i\alpha}(\lambda t')H^{(1)}_{i\alpha}(\lambda t) \right).\nonumber
\eea
This is same as the one we get for the Minkowski vacuum.
Likewise, in $4+1$ dimensions, in the Minkowski vacuum $G(x,x')$ is
\bea
G(x,x^\prime) = \frac{\pi(zz^\prime)^2}{4tt^\prime}\int  \lambda d\lambda d\alpha   
J_\nu(\lambda z)J_\nu(\lambda z^\prime)
\left( H^{(2)}_{i\alpha}(\lambda t)
H^{(1)}_{i\alpha}(\lambda t')-H^{(2)}_{i\alpha}(\lambda t')
H^{(1)}_{i\alpha}(\lambda t) \right) \frac{\alpha \sin \alpha\xi}{2\pi^2\sinh\xi},\nonumber
\eea
and for conformal vacuum, the corresponding expression is
\bea
G(x,x^\prime) = \frac{\pi(zz^\prime)^2}{4tt^\prime}\int  
\lambda d\lambda d\alpha   J_\nu(\lambda z)J_\nu(\lambda z^\prime)
\left( J_{-i\alpha}(\lambda t)J_{i\alpha}(\lambda t')-
J_{i\alpha}(\lambda t)J_{-i\alpha}(\lambda t') \right) 
\frac{\alpha \sin \alpha\gamma}{2\pi^2\sinh\pi\alpha\sinh\gamma}.\nonumber
\eea
Using (\ref{bessel}), we immediately see that the last two expressions coincide. 
This is not surprising. $G(x,x')$, also known as the
Pauli-Jordan function, arises from a commutator of the fields and 
is supposed to be a state independent function.

We can also give a general check that the retarded propagators or 
the Pauli-Jordan functions are same in both the vacua. We start with
\be
G^{+}(x,y) = \sum_{i} \phi_{i}^{*}(x) \phi_{i}(y)  \ \ , \ \   G^{-}(x,y)= \sum_{i} \phi_{i}^{*}(y) 
\phi_{i}(x)\nonumber
\ee
Here $\phi_i$ are modes calculated in one vacuum (e.g. Minkowski vacuum) 
and summation represents the integration over all continuous indices
and the sum over the discrete indices. Using the Bogoliubov coefficients, 
we can express one set of modes $\phi_i$  in terms of other set $\tilde{\phi}$ as
\be
\phi_i (y) = \sum_j \left( A^{*}_{ji} \tilde \phi_j (y) -B_{ji} \tilde\phi_{j}^{*}(y) \right), \  \
\phi^{*}_{i} (x) = \sum_k \left( A_{ki} \tilde\phi^{*}_{k} (x) -B^{*}_{ki} \tilde\phi_{k}(x) \right), \label{bogo}
\ee
where Bogoliubov coefficients satisfy \cite{Birrell:1982ix}

\be
\sum_k \left(A_{ik}A^{*}_{jk} -B_{ik}B^{*}_{jk}\right) = \delta_{ij}, 
\  \  \sum_k \left(A_{ik}B_{jk} -B_{ik}A_{jk}\right) = 0. \label{bogoprop}
\ee
It is easy to check that Bogoliubov coefficients in our case indeed satisfy these identities. \footnote{
In $3+1$ dimensions, we see that the  transformation equation is
\be
\tilde\phi_{\lambda\alpha n}= \sum_{\lambda'\alpha' n'}\left( A_{\lambda\alpha n \lambda'\alpha' n'}
\phi_{\lambda'\alpha' n'} +B_{\lambda\alpha n \lambda'\alpha' n'}\phi^{*}_{\lambda'\alpha' n'}\right).\nonumber
\ee
Using \ref {solutionseq} and \ref{solutionseq2} we see that the Bogoliubov coefficients are
\be
A_{\lambda\alpha n \lambda'\alpha' n'}= \delta_{\lambda\lambda'}\delta_{nn'}\delta_{\alpha\alpha'}c(\alpha) \ \  ,
 \  \ B_{\lambda\alpha n \lambda'\alpha' n'}= \delta_{\lambda\lambda'}\delta_{n,-n'}\delta_{\alpha\alpha'}
d(\alpha).\nonumber
\ee
Using the relationship between Hankel and Bessel functions, along with 
$H^{(2)}_{i\alpha'} = e^{-\pi\alpha'}H^{(2)}_{-i\alpha'}$, we get
\be
c(\alpha) = \frac{e^{\frac{\pi \alpha}{2}}}{\sqrt{2\sinh\pi\alpha}},  
\  \ d(\alpha)= -\frac{e^{-\frac{\pi \alpha}{2}}}{\sqrt{2\sinh\pi\alpha}}.\nonumber
\ee
}
Further, putting the expansion (\ref{bogo}) in the Wightman functions, we get
\be
G_+(x,y) = \sum_{i,j,k} \left( A_{ki}A^{*}_{ji}\tilde\phi_{k}^{*}(x) 
\tilde\phi_{j}(y) + B_{ji}B^{*}_{ki}\tilde\phi_{j}^{*}(y) 
\tilde\phi_{k}(x) -
A_{ki}B_{ji}\tilde\phi_{k}^{*}(x) \tilde\phi^{*}_{j}(y) -
B^{*}_{ki}A^{*}_{ji}\tilde\phi_{k}(x) \tilde\phi_{j}(y) \right).\nn
\ee
and
\be
G_-(x,y) = \sum_{i,j,k} \left( A_{ki}A^{*}_{ji}\tilde\phi_{k}^{*}(y) 
\tilde\phi_{j}(x) + B_{ji}B^{*}_{ki}\tilde\phi_{j}^{*}(x) \tilde\phi_{k}(y) -
A_{ki}B_{ji}\tilde\phi_{k}^{*}(y) \tilde\phi^{*}_{j}(x) -B^{*}_{ki}A^{*}_{ji}
\tilde\phi_{k}(y) \tilde\phi_{j}(x) \right).\nn
\ee
Subtracting the above two expressions and using (\ref{bogoprop}), we  get the Pauli-Jordan function $G(x,y)$.
\be
G (x,y)=  \sum_{i} \left( \phi_{i}^{*}(x) \phi_{i}(y)  -\phi_{i}^{*}(y) \phi_{i}(x) \right) = \sum_{i}
\left( \tilde\phi_{i}^{*}(x) \tilde\phi_{i}(y)  - \tilde\phi_{i}^{*}(y) \tilde\phi_{i}(x) \right).\nonumber
\ee

Equipped now with these results, in the next section, we analyse scalar field theory on the AdS-Milne spacetime
containing a conical defect. Here this defect runs all the way to the boundary. Our primary aim would be  to find 
boundary two point correlator of operators dual to the bulk field in a closed form. 

\subsection{AdS-Milne in other vacua}

So far we have focused on the adiabatic and the conformal vacua. Recently, in 
\cite{Akhmedov:2021agm}, a more general class of vacua, similar to the 
alpha-vacua of the de Sitter spacetime, was considered. We end this section with a computation of the Wightman 
function in these vacua. Our result is similar to the one of \cite{Akhmedov:2021agm}, namely, the Wightman function 
picks up a dependence on the coordinates in a non-Poincare invariant manner.

To proceed, we  start with the general solution of (\ref{kgeqn})  
\begin{eqnarray}
\phi_{\lambda \alpha n} (z, t, r, \theta) = \Big[z^{\frac{3}{2}} J_{\nu}(\lambda z) \Big]
\Big[ \frac{  C_{\lambda \alpha n} 
H_{i\alpha}^{(2)}(\lambda t) + \tilde C_{\lambda \alpha n} H_{i\alpha}^{(1)}(\lambda t')}{\sqrt t}\Big]
P^{-n}_{i\alpha-\frac{1}{2}}({\rm
cosh}r)e^{in\theta}.
\label {solutionalpha}
\end{eqnarray}
The normalization condition (\ref{normalization}) for the above solution gives 
\be
|C_{\lambda \alpha n} |^2 e^{-\pi \alpha} - |\tilde C_{\lambda \alpha n} |^2 e^{\pi \alpha} 
= - \frac{\alpha\lambda}{8\pi} |\Gamma( n + \frac{1}{2} + i\alpha)|^2 \sinh \pi\alpha. \label{normconstant}
\ee
It can be checked that the choice 
\be
C_{\lambda\alpha n}= \frac{i}{2}\sqrt{\frac{\alpha\lambda}{4\pi}}e^{\pi\alpha} 
\Gamma(i\alpha + \frac{1}{2} + n) \ \ , 
\  \tilde C_{\lambda\alpha n}= \frac{i}{2}\sqrt{\frac{\alpha\lambda}{4\pi}}e^{-\pi\alpha} 
\Gamma(i\alpha + \frac{1}{2} + n) \label{normconf}
\ee
gives the solution (\ref{phiconformal}) in the conformal vacuum 
with the correct normalization factor. Related to the adiabatic modes 
is a two parameter family of modes, called alpha-modes.  We can write the normalization 
constants of these in terms of a two-parameter family of constants labelled by $\rho , \sigma$ as
\begin{eqnarray}
&&C_{\lambda\alpha n}= i\cosh \rho \sqrt{\frac{\alpha\lambda \sinh \pi\alpha}{8\pi}}e^{\frac{\pi\alpha}{2}} 
\Gamma(i\alpha + \frac{1}{2} + n) \nonumber \\
&&\tilde C_{\lambda\alpha n}= i\sinh \rho \ e^{i\sigma} \sqrt{\frac{\alpha\lambda \sinh\pi\alpha}{8\pi}}
e^{-\frac{\pi\alpha}{2}} \Gamma(i\alpha + \frac{1}{2} + n)\sqrt{\sinh\pi\alpha} \label{alphanorm}
\end{eqnarray}
For $\rho , \sigma$ set to zero, we get back  the adiabatic modes 
(\ref{solutionseq}) with the correct normalization constant. The above normalization 
constants satisfy the relation (\ref{normconstant}) . Since $\rho , \sigma$ are constants, 
independent of $\alpha$, we can see that this family does not contain the conformal modes 
given by (\ref{normconf}) . We can calculate the Wightman function for these alpha-modes by 
putting (\ref{solutionalpha}) in the relation (\ref{wightman}) along with the normalization constants 
in (\ref{alphanorm}) and using the appropriate alpha vacua states. 
The Wightman function for the field in the alpha vacua then comes out as
\begin{eqnarray}
G_+(x,x^\prime)&=&   -\frac{\cosh^2 \rho \ Q_{\nu - 1/2}^1(u_1)}{4\pi^2 (u_{1}^2 -1)^{\frac{1}{2}}}  
- \frac{\sinh^2 \rho\  Q_{\nu - 1/2}^1(u_2)}{4\pi^2 (u_{2}^2 -1)^{\frac{1}{2}}}  + 
i \sinh \rho\cosh\rho\frac{ e^{-i\sigma}\ Q_{\nu - 1/2}^1(u_3)}{ 4\pi^2(u_{3}^2 -1)^{\frac{1}{2}}} \nonumber  \\ 
&+&  \frac{i \sinh \rho\cosh\rho ~e^{i\sigma}\  Q_{\nu - 1/2}^1(u_4)}{ 4\pi^2(u_{4}^2 -1)^{\frac{1}{2}}}, \nonumber 
\end{eqnarray}
where 
\begin{eqnarray}
u_1 &=& \frac{1}{2 z z'} \Big(-t^2 - t'^2 + 2 t t' ~{\rm cosh}\chi + z^2 + z'^2 + i\epsilon ~{\rm sgn}(t-t')\Big), \nonumber\\
u_2 &=& \frac{1}{2 z z'} \Big(-t^2 - t'^2 + 2 t t' ~{\rm cosh}\chi + z^2 + z'^2 - i\epsilon ~{\rm sgn}(t-t')\Big), \nonumber\\
u_3 &=& \frac{1}{2 z z'} \Big(-t^2 - t'^2 - 2 t t' ~{\rm cosh}\chi + z^2 + z'^2 + i\epsilon \Big), \nonumber\\
u _4 &=& \frac{1}{2 z z'} \Big(-t^2 - t'^2 - 2 t t' ~{\rm cosh}\chi + z^2 + z'^2 - i\epsilon )\Big).\nonumber
\end{eqnarray}
First term is just  $\cosh^2 \rho$ times the Wightman function in the adiabatic vacuum (\ref{bulkwightman}). 
Second term is $\sinh^2 \rho$ times the Wightman function in the adiabatic vacuum, 
but with $t \leftrightarrow t'$. Other two terms are not Poincare invariant 
and are analogous to the dependence on antipodal distance as discussed 
in \cite{Akhmedov:2021agm}. 
 
 \section{\bf AdS-Milne with a conical defect}

AdS-Milne background in the presence of a conical defect has the same form as before except that the
the angular coordinate has a different periodicity. In $3+1$ dimensions, the metric is given by
\begin{equation}
ds^2 = \frac{1}{z^2}\Big[dz^2 - dt^2 + t^2 dr^2 + t^2 ~{\rm sinh^2r}~d\theta^2\Big],
\label{adsmilnecone}
\end{equation}
where $0\le \theta <2\pi/q$. We will take $q > 1$.

The above geometry may arise in the presence of an 
infinite string in a four dimensional AdS-Milne spacetime. This can be seen by closely following the treatment 
of \cite{Bayona:2010sd}. Consider the Nambu-Goto action 
\begin{equation}
S_{NG} = -\mu \int d\sigma_0 d\sigma_1 \sqrt{-det P[g_{ab}]}, 
\ \ P[g_{ab}]= g_{\mu\nu}(X) \frac{\partial X^\mu (\sigma)}{\partial \sigma^a }
\frac{\partial X^\nu (\sigma)}{\partial \sigma^b }.\end{equation}
Here $\mu$ is the tension associated with the string.
We choose a gauge such that $\sigma^0 ,\sigma^1 = t,z$ and consider the string 
extended along the $z$ direction. Embedding coordinates will then be $X^\mu (t,z) = \left(t,z, r(t,z),\theta(t,z) \right)$. 
In general, we can vary Nambu-Goto action with respect to embedding coordinates to get the equations of motion. 
But we are interested in a particular solution corresponding to string at the origin of $(r,\theta)$ plane, 
represented by a delta-function source $\delta(r)$. 

Corresponding Nambu-Goto action and its variation with respect to the metric is given by 
\begin{equation}
S= -\frac{\mu}{\pi} \int dt dz dr d\theta \sqrt{-g_{tt}g_{zz}}\delta(r),
\ \  \delta S = \frac{\mu}{2\pi} \int dt dz dr d\theta \left(\frac{g_{zz} 
\delta g_{tt} + g_{tt}\delta g_{zz}}{\sqrt{-g_{tt}g_{zz}}}\right)\delta(r).
\end{equation}

Comparing this with the general relation between metric variation and stress tensor
\begin{equation}
\delta S = \frac{1}{2}\int  dt dz dr d\theta \sqrt{-g} T^{\mu\nu}\delta g_{\mu\nu},
\end{equation}
we can get read off the stress tensor corresponding to the string configuration. Equivalently, we can do a coordinate transformation of the result given in \cite{Bayona:2010sd}. We get 
\begin{equation}
T^{\mu\nu}= -\frac{\mu\delta(r)t}{\pi \sqrt{g_{rr} g_{\theta\theta}}}
\begin{pmatrix}
g_{tt}^{-1} & 0 & 0 & 0 \\
0 & 0 & 0 & 0 \\
0 & 0 & 0 & 0 \\
0 & 0 & 0 & g_{zz}^{-1}
\end{pmatrix}.
\end{equation}
Taking this as the stress tensor for the string, we solve the coupled Einstein-Hilbert and the Nambu-Goto action
\begin{equation}
 S= \frac{1}{16 \pi G_4}\int d^4 x \sqrt{-g}\left( R+ \frac{6}{L^2}\right) -\mu \int d\sigma_0 d\sigma_1 \sqrt{-det P[g_{ab}]}.
  \end{equation}
Variation of the above leads to 
  \begin{equation}
  R^{\mu\nu} -\frac{R}{2} g^{\mu\nu} -\frac{3}{L^2} g^{\mu\nu} = 8\pi G_4 T^{\mu\nu}.
  \end{equation}
Solving this system of equations we get the metric
 \begin{equation}
ds^2 = \frac{1}{z^2}\Big[dz^2 - dt^2 + t^2 dr^2 + t^2 ~{\rm sinh^2r}~d\theta^2\Big],
\end{equation}
with $0 \le t, r \le \infty$ and $0 \le \theta \le 2\pi(1-4\mu G_4)$. This is same as (\ref{adsmilnecone}) once we identify
$q = (1-4\mu G_4)$.

We now turn to the  scalar propagator on this geometry in the Minkowski vacuum. In the following we will primarily restrict 
$q$ to be an integer.
The scalar field now needs to satisfy
$\phi(z, t, r, \theta) = \phi(z, t, r, \theta + 2\pi/q)$. Solving the equation of motion, we find
\begin{eqnarray}
\phi(z, t, r, \theta) = \frac{i}{2}\sqrt{\frac{q \alpha\lambda ~{\rm 
sinh}\pi\alpha}{2\pi}}\Gamma(i\alpha + 
\frac{1}{2} + q n) e^{\pi\alpha/2} 
\Big[z^{\frac{3}{2}} J_{\nu}(\lambda z) \Big]
\Big[ \frac{H_{i\alpha}^{(2)}(\lambda t)}{\sqrt t}\Big]
P^{-qn}_{i\alpha-\frac{1}{2}}({\rm
cosh}r)e^{iqn\theta}.\nonumber
\end{eqnarray}
Therefore, now the Wightman function takes the form 
\begin{eqnarray}
G^q_+(x, x') &=& \frac{q(zz')^{\frac{3}{2}}}{8\pi\sqrt{tt'}}\int \sum_{n=-\infty}^{\infty}\lambda~d\lambda 
~\alpha 
~d\alpha ~
   e^{\pi\alpha}~{\rm sinh}(\pi\alpha) J_\nu(\lambda z) J_\nu(\lambda z') H_{i\alpha}^{(2)}(\lambda t)
H_{-i\alpha}^{(1)}(\lambda t')\nonumber\\
&\times& |\Gamma(i\alpha + 1/2 + qn)|^2 P_{i\alpha -1/2}^{-qn}({\rm cosh}r) P_{i\alpha -1/2}^{-qn}({\rm 
cosh}r')e^{iqn(\theta - \theta')}.\nonumber
\end{eqnarray}
Further, using 
\begin{eqnarray}
H_{i\alpha}^{(2)}(\lambda t) H_{-i\alpha}^{(1)}(\lambda t')= \frac{4}{\pi^2} e^{-\pi\alpha} K_{i\alpha}(i\lambda t) 
K_{i\alpha}(-i\lambda t'),\nonumber
\end{eqnarray}
we can rewrite 
\begin{eqnarray}
G^q_+(x,x') &=& \frac{q(zz')^{\frac{3}{2}}}{2\pi^3\sqrt{tt'}}\int \sum_{n=-\infty}^{\infty}\lambda~d\lambda ~\alpha
~d\alpha ~
   e^{\pi\alpha}~{\rm sinh}(\pi\alpha) J_\nu(\lambda z) J_\nu(\lambda z') K_{i\alpha}(i\lambda t)
K_{i\alpha}(-i\lambda t')\nonumber\\
&\times& |\Gamma(i\alpha + 1/2 + qn)|^2 P_{i\alpha -1/2}^{-qn}({\rm cosh}r) P_{i\alpha -1/2}^{-qn}({\rm
cosh}r')e^{iqn(\theta - \theta')}.
\label{interexp}
\end{eqnarray}
Using (\ref{appen4}), we can simplify (\ref{interexp}) as,
\begin{eqnarray}
G^q_+(x,x') &=& \frac{q(zz')^{\frac{3}{2}}}{4\pi\sqrt{2\pi tt'}}\int_0^\infty \lambda 
~d\lambda\sum_{n=-\infty}^{\infty} 
\int_0^\infty dx ~x^{-1/2}
J_\nu(\lambda z) J_\nu(\lambda z') e^{\Big( \frac{t^2 + t'^2}{2tt'} x - \frac{\lambda^2 tt'}{2x}\Big)}
\nonumber\\
&\times& I_{|n|q}(x~{\rm sinh}r~{\rm sinh}r')e^{-x~{\rm cosh}r~{\rm cosh}r'}e^{iqn(\theta - \theta')}.
\label{bbbulk}
\end{eqnarray}
Now the $\lambda$ integral can be performed using \ref{lambdaintegral} to get, after a little algebra\footnote{Some of our manipulations here are similar to that of 
\cite{deMello:2011ji}.},
\begin{eqnarray}
G^q_+(x,x') &=& \frac{q (zz'/tt')^{\frac{3}{2}}}{4\pi \sqrt{2\pi }}\sum_{n=-\infty}^\infty e^{iqn(\theta - \theta')}
\int_0^\infty dx \sqrt{x}  I_{|n|q}(x~{\rm sinh}r~{\rm sinh}r') 
I_\nu\Big(\frac{zz'x}{tt'}\Big)\nonumber\\
&\times& ~{\rm exp}\Big[ -\frac{x}{2tt'} (z^2 + z'^2 -t^2 - t'^2 + 2 tt'~{\rm cosh}r~{\rm 
cosh}r')\Big]\nonumber\\
&=& q\Big(\frac{zz'}{4\pi}\Big)^{\frac{3}{2}} \ \int_0^\infty \frac{ds}{s^4} e^{-\frac { {\cal{V}}^2 }{4s^2}}
S_q\Big( \frac{tt'~{\rm sinh}r~{\rm sinh}r'}{2s^2}\Big) I_\nu\Big(\frac{zz'}{2s^2}\Big).
\label{detail1}
\end{eqnarray}
In arriving at the last equation, we have used $s^2 = \frac{tt'}{2x}$ and 
\begin{eqnarray}
S_q\Big( \frac{tt'~{\rm sinh}r~{\rm sinh}r'}{2s^2}\Big)  &=& \sum_{n=-\infty}^\infty e^{inq(\theta-\theta')} 
I_{|n|q} \Big( \frac{tt' ~{\rm sinh}r~{\rm sinh}r'}{2s^2}\Big)\nonumber\\
&=& 2\sum_{n = 0}^{\infty~'} ~{\rm cos}[nq(\theta -\theta')] I_{nq}\Big(\frac{tt' ~{\rm sinh}r~{\rm 
sinh}r'}{2s^2}\Big)\label{sssum}\\
&=& \frac{1}{q}\sum_{k=0}^\infty ~{\rm exp}\Big[ \Big( \frac{tt' ~{\rm sinh}r~{\rm sinh}r'}{2s^2}\Big)~{\rm 
cos}\Big(\theta -\theta' - \frac{2k\pi}{q}\Big)\Big].
\label{ssum}
\end{eqnarray}
In (\ref{detail1}),
\begin{equation}
{\cal{V}}^2 = z^2 + z'^2 - t^2 - t'^2 + 2tt'~{\rm cosh}r~{\rm cosh}r'.\nonumber
\end{equation}
The prime on the sum in equation (\ref{sssum}) means that the contribution from $n =0$ term should be halved.
Since we are working with integer values of $q$, the $k$ sum will take integer values from $0$ to $q-1$.
We can now  carry out the integration over $s$ using \ref{xintegral} to get the final expression of the Wightman function 
\begin{equation}
G^q_+(x,x') = -\frac{1}{4\pi^2} \sum_{k=0}^{q-1} \frac{Q_{\nu -1/2}^1(u_k)}{\sqrt{u_k^2 
-1}}.
\label{renW}
\end{equation}
Here $u_k$ is defined as
\begin{eqnarray}
u_k &=& \frac{1}{2zz'} (z^2 + z'^2 - t^2 - t'^2 + 2tt' ~{\rm cosh}\chi_{kq}),
\label{uk}
\end{eqnarray}
and
\begin{eqnarray}
{\rm cosh}\chi_{k,q} &=& ~{\rm cosh}r~{\rm cosh}r' - ~{\rm sinh}r~{\rm sinh}r'
~{\rm cos}(\theta - \theta' - 2\pi 
k/q).\nonumber
\end{eqnarray}

The expression for the Wightman function is reminiscent of the method of images. In the coincident limit, $G^q_+(x,x')$ diverges. This divergence comes from the $k =0$ term of the sum.
We can define a renormalized function subtracting this contribution and write
\begin{equation}
G^q_{+R}(x,x') = -\frac{1}{4\pi^2} \sum_{k=1}^{q-1} \frac{Q_{\nu -1/2}^1(u_k)}{\sqrt{u_k^2
-1}}.
\label{renormW}
\end{equation}

\bigskip

\noindent{\underbar{Boundary correlator}}

\bigskip

The boundary correlator is constructed as before
\begin{eqnarray}
\langle \Psi_q| {\cal O}(t,r,\theta) {\cal O}(t', r', \theta')|\Psi_q\rangle &=& ~ \lim\limits_{z,z' \rightarrow 0} 
(zz')^{-(\nu +3/2)} G^q_+(x,x') \nonumber\\
&=& \sum_{k=0}^{q-1} \frac{C_\Delta}{(-t^2 - t'^2 + 2tt'~{\rm cosh}\chi_{k,q})^\Delta},
\label{boundaryc}
\end{eqnarray}
where $C_\Delta$ is a constant defined in (\ref{cdelta}).

The summation over $k$ can be performed and the final result comes out as
\begin{eqnarray}
\langle \Psi_q|{\cal O}(t,r,\theta) {\cal O}(t', r', \theta')|\Psi_q\rangle =
\frac{(-1)^{\Delta -1}C_\Delta}{\Gamma(\Delta ) (2t t' ~{\rm sinh}r ~{\rm sinh}r')^\Delta} 
\Big(\frac{\partial^{\Delta -1}}{\partial
\gamma^{\Delta-1}}\Big)I_1(q, \gamma,
\theta
-
\theta'),
\label{lasteq}
\end{eqnarray}
where,
\begin{eqnarray}
I_1(q, \gamma, \theta-\theta') =
\Big[ \frac{ q[(\gamma + \sqrt{\gamma^2 -1})^{2q} - 1]}
{\sqrt{\gamma^2 -1} [ 1 + (\gamma + \sqrt{\gamma^2 -1})^{2q} - ~2 {\rm cos}q(\theta - \theta')(\gamma +
\sqrt{\gamma^2 -1})^q ]}\Big],\nonumber
\end{eqnarray}   
and
\begin{eqnarray}
\gamma = \frac{-t^2 - t'^2 + 2 tt' ~{\rm cosh}r~{\rm cosh}r'}{2tt'~{\rm sinh}r~{\rm sinh}r'}.\nonumber
\end{eqnarray}
The details are provided in the appendix.

\bigskip

\noindent{\underbar{Twisted scalar}}

\bigskip

Having come thus far, we end this section with a study of the correlators 
involving twisted fields. These fields satisfy the same equations as the untwisted scalars but differ 
in their boundary conditions.  The quasi-periodic boundary condition that the twisted scalars obey is
given by
\begin{equation}
\phi(z, t, r, \theta + \frac{2\pi}{q}) = e^{-2\pi i\beta} \phi(z, t, r, \theta),
\label{tperiod}
\end{equation}
where $0 \le \beta \le 1$. Such twisted scalar fields arise, for example, when we consider a charged scalar field in AdS in the presence of 
a cosmic string carrying internal magnetic flux. As is well known (see, for example \cite{euclideanconical} and references therein), the corresponding gauge field component can be eliminated by a gauge transformation and then one is left with a scalar with twisted boundary conditions. For earlier studies along this direction, see for example \cite{Souradeep:1992ia}, \cite{Farias}.

Normalized solution of the equation of motion is now
\begin{eqnarray}
\phi(z, t, r, \theta) = C_{\lambda \alpha n} z^{\frac{3}{2}} J_\nu(\lambda z) \frac{H^{2}_{i\alpha}(\lambda t)}{\sqrt
t} P^{q(n-\beta)}_{i\alpha -1/2} ({\rm cosh}r) e^{i q (n - \beta)\theta},\nonumber
\end{eqnarray}
where
\begin{eqnarray}
C_{\lambda \alpha n} = \frac{ie^{\pi \alpha/2}}{2} \sqrt{ \frac{q \alpha~{\rm sinh}(\pi\alpha)}{2\pi} }
\Gamma[q(n-\beta) + i\alpha
+ 1/2].\nonumber
\end{eqnarray}
Calculation is quite similar to the previous case of $\beta=0$ and as before the Wightman function takes the form
\begin{equation}
W_{\beta}^q(x,x') = q\Big(\frac{zz'}{4\pi}\Big)^{\frac{3}{2}}
\int_0^\infty \frac{ds}{s^4} e^{-\frac{{\cal{V}}^2}{4 s^2}} S_{q\beta}\Big(\frac{tt'~{\rm sinh}r~{\rm
sinh}r'}{2s^2}\Big)
I_\nu\Big(\frac{zz'}{2s^2}\Big),
\label{twistedW}
\end{equation}
where now
\begin{eqnarray}
 S_{q\beta}\Big(\frac{tt'~{\rm sinh}r~{\rm sinh}r'}{2s^2}\Big) = \sum_{n = -\infty}^{\infty}
e^{i(n-\beta)q (\theta - \theta')} I_{|n-\beta|q}\Big(\frac{tt'~{\rm sinh}r~{\rm sinh}r'}{2s^2}\Big),\nonumber
\end{eqnarray}
and
\begin{eqnarray}
{\cal{V}}^2 = z^2 + z'^2 - t^2 - t'^2 + 2 tt'{\rm cosh}r~{\rm cosh}r'.\nonumber
\end{eqnarray}
We can now use the following relation to carry out the $s$ integration \cite{deMello:2014ksa}:
\begin{eqnarray}
&&\sum_{n =  -\infty}^{\infty} e^{iq(n-\beta)(\theta - \theta')}I_{|n-\beta|q}\Big(\frac{tt'~{\rm sinh}r~{\rm
sinh}r'}{2s^2}\Big) = \nonumber\\
&&~~~~~~~\sum_{n}\Big[\frac{1}{q} e^{[{tt'~{\rm sinh}r~{\rm sinh}r'~{\rm cos}(2\pi n/q - \theta +
\theta')}]/{2s^2}}
e^{i\beta(2\pi n - q\theta + q \theta')}\nonumber\\
&&~~~~~~~- \frac{1}{2\pi i} \sum_{j=+, -} j e^{j i \pi q\beta} \int_0^\infty dy
e^{-{tt'~{\rm sinh}r~{\rm sinh}r'~{\rm cosh}y}/{2s^2}}f(y)\Big] e^{-iq\beta(\theta - \theta')},
\label{bezerra}
\end{eqnarray}
where
\begin{eqnarray}
f(y) = \frac{{\rm cosh}[qy(1 - \beta)]~-~{\rm cosh}(q\beta y)e^{-iq(\theta-\theta' + j\pi)} }
{{\rm cosh}(qy)~-~{\rm cos}[q(\theta-\theta' + j\pi)]}.\nonumber
\end{eqnarray}
Here the sum over $n$ runs as
\begin{equation}
-\frac{q}{2} + \frac{(\theta - \theta')q}{2\pi} \le n \le + \frac{q}{2} + \frac{(\theta - \theta')q}{2\pi}.
\end{equation}  
Once substituted (\ref{bezerra}) into (\ref{twistedW}), we can integrate over $s$ using 
\begin{eqnarray}
&&\int_0^\infty \frac{ds}{s^4} e^{-[\frac{{\cal{V}}^2}{2} - tt'~{\rm sinh}r~{\rm sinh}r'~{\rm cos}(2\pi n/q -
\theta
+
\theta')]/(2s^2)} I_\nu\Big(\frac{zz'}{2 s^2}
\Big) =\nonumber\\
&&~~~~~~~- \frac{2}{(zz')^{3/2} \sqrt{\pi}\sqrt{u_{n}^2 -1}} Q^1_{\nu - 1/2} (u_{n}), \nonumber
\end{eqnarray}
where $u_n$ is defined as in (\ref{uk}), to finally arrive at
\begin{eqnarray}
W_{\beta}^q(x, x') &=& -\frac{1}{(2\pi)^2} \sum_n e^{-2i\pi\beta n} \frac{Q_{\nu
-1/2}^1(u_n)}{\sqrt{u_n^2-1}}\nonumber\\
&-& \frac{iq}{(2\pi)^3} e^{-iq\beta(\theta -\theta')} \sum_{j=+,-} j e^{ji\pi q\beta}\int_0^\infty dy
f(y) \frac{Q_{\nu-1/2}^1(u_y)}{\sqrt{u_y^2-1}},
\end{eqnarray}
where $u_y$ has the same structure of (\ref{uk}) with ${\rm cosh}\chi_{kq}$ replaced by ${\rm cosh}y$.
This is the general form of the Wightman function. 

When $q$ and $q\beta$ are both integers, the second term vanishes and we get
\begin{eqnarray}
W_{\beta}^q(x, x') &=& -\frac{1}{(2\pi)^2} \sum_n e^{-2i\beta\pi n } \frac{Q_{\nu
-1/2}^1(u_n)}{\sqrt{u_n^2-1}}.
\label{twistedco}
\end{eqnarray}

The twisted bulk scalar is dual to a twisted scalar scalar primary ${\cal{O}}_\beta$ of dimension 
$\Delta$ with the same periodicity (\ref{tperiod}) along $\theta$ inherited from $\phi$. 
The boundary correlator is therefore,
\begin{equation}
\langle \Psi_{\beta q}| {\cal O}_\beta(t,r,\theta) {\cal O}_\beta(t', r', \theta')|\Psi_{\beta q}\rangle =
\sum_{n} \frac{C_\Delta e^{-2i\beta\pi n}}{(-t^2 - t'^2 + 2tt'~{\rm cosh}\chi_{n,q})^\Delta},
\end{equation}
where the constant $C_\Delta$ has been defined earlier. This series can be summed to get

\begin{eqnarray}
\langle \Psi_{\beta q}| {\cal O}_\beta(t,r,\theta) {\cal O}_\beta(t', r', \theta')|\Psi_{\beta q}\rangle = 
\frac{(-1)^{\Delta -1}C_\Delta}{\Gamma(\Delta ) (2t t' ~{\rm sinh}r ~{\rm sinh}r')^\Delta} 
\Big(\frac{\partial^{\Delta -1}}{\partial
\gamma^{\Delta-1}}\Big) J^{q}_{\beta}( \gamma,
\theta
-
\theta'),
\label{twistedsum}
\end{eqnarray}
where, $J^{q}_{ \beta}(\gamma, \theta-\theta') =J$ is given by 

\begin{eqnarray}
J =
 \frac{ q e^{-iq\beta(\theta -\theta')} [(\gamma + \sqrt{\gamma^2 -1})^{2q} - (\gamma + \sqrt{\gamma^2 -1})^{q\beta} + 2e^{iq(\theta-\theta')}
 (\gamma + \sqrt{\gamma^2 -1})^{q} \sinh(q\beta \ln(\gamma + \sqrt{\gamma^2 -1})]}
{\sqrt{\gamma^2 -1} [ 1 + (\gamma + \sqrt{\gamma^2 -1})^{2q} - ~2 {\rm cos}q(\theta - \theta')(\gamma +
\sqrt{\gamma^2 -1})^q ]]},\nonumber
\end{eqnarray}   
and
\begin{eqnarray}
\gamma = \frac{-t^2 - t'^2 + 2 tt' ~{\rm cosh}r~{\rm cosh}r'}{2tt'~{\rm sinh}r~{\rm sinh}r'}.\nonumber
\end{eqnarray}
The details are provided in the appendix.

\bigskip

\section{Conclusions}

We have initiated a holographic study of field theory on the time dependent background with a conical defect.
In this work, our focus has been on the Milne spacetime to which, in the absence of cosmological constant, 
 any hyperbolic FRW metric flows to at late times. When the Milne vacuum is chosen to be the adiabatic one,
we are able to compute the two point correlators of operators which are dual to the massive scalars in the
bulk AdS-Milne background with defect. We find, for both twisted and untwisted operators, the 
correlators can be represented as the sum over images in the covering space. This sum can be
carried out explicitly to write the results in compact forms. If we restrict ourselves to the adiabatic vacuum,
our computations suggest that the field theory defined on a part of the boundary of AdS is dual to a subregion 
in the AdS bulk. Though it may not be entirely obvious, there are indication that the subregion duality 
may hold in general \cite{CHM}-\cite{RT}.

Evidently, our exploration is incomplete. First of all, Milne spacetime offers another natural
vacuum known as the conformal vacuum. We have not been able get a closed form expression of correlators in this
vacuum. Further, the renormalized stress tensors, which one calculates from the two
point correlators, for the conformal vacuum are known to be non-trivial in Milne spacetime. 
This remains to be computed in the presence of the defect in this holographic framework.

In case of the AdS-Rindler foliation of AdS, interesting progress has been made 
in understanding the relationship between entanglement 
and spacetime \cite{Czech:2012be}. Since AdS-Milne is another foliation of AdS, albeit, time-dependent, 
it will be instructive to study the relationship between spacetime and
entanglement structure for the time-dependent situation. Apart from these considerations, our results also connect with
 recent studies of conical defects in $AdS_3$ , for example, \cite{Berenstein:2022ico}. We believe our results will also be useful
  in the study of entanglement entropy in time-dependent situations in a  holographic context. 

We are currently exploring some of these issues and hope to report on them in the near future.

\bigskip

\noindent {\bf Acknowledgements:} We would like to especially thank Souvik Banerjee for the initial collaboration  and many insightful discussions throughout the project. 

\bigskip

\bigskip

\section{Appendix}

\bigskip

\bigskip
 
\noindent\underbar{$4+1$ dimensional AdS-Milne}

\bigskip

\bigskip

Since  mode expansions of a massive scalar is sensitive to the spacetime dimensions, it is instructive to
carry out a computation in $4+1$ dimensions. The metric is given in ({\ref{milnefour}). 
The relevant solution of the equation of motion for the Minkowski vacuum turns out to be
\bea
\phi(z,t,r,\theta,\phi) =
C_{\alpha lm}  z^2J_\nu(\lambda z)\frac{H_{i\alpha}^{(2)}(\lambda t)}{t}  Y_{\alpha lm}(r,\theta,\phi),\nn
\eea
where $\nu = \sqrt{4+m^2}$ and we have defined \cite{5}- \cite{6}
\bea
Y_{\alpha l m}(r,\theta,\phi) =
\frac{\Gamma(i\alpha+l+1)}{\Gamma(i\alpha+1)}\frac{\alpha}{\sqrt{\sinh r}}P_{i\alpha-1/2}^{-l-1/2}(\cosh r)
Y_{l m}(\theta,\phi)\nn
\eea
The spherical harmonics $Y_{lmp}(\chi,\theta,\phi)$, for $0<p<\infty$, form a complete orthonormal
set of square integrable functions on the unit hyperboloid \cite{5}-\cite{6}
\bea
\int_0^\infty d\chi\sinh^2\chi\int d\Omega Y_{lmp}(\chi,\Omega)Y_{l^\prime m^\prime p^\prime}^*(\chi, \Omega) =
\delta(p-p^\prime)\delta_{ll^\prime}\delta_{m m^\prime}.\nn\label{N5}
\eea
where the harmonic functions satisfy
\bea
\int_0^\infty d\chi\sinh\chi P_{ip-1/2}^{-l-1/2}(\cosh\chi)P_{ip\prime-1/2}^{-l^\prime-1/2}(\cosh\chi) =
\frac{|\Gamma(ip)|^2}{|\Gamma(ip+l+1)|^2}\delta(p-p^\prime).\nn
\eea
The normalization constant $C_{\alpha l m}$ turns out to be
\bea
C_{\alpha l m} = i\frac{\sqrt{\pi\lambda}}{2}e^{\pi\alpha/2}.\nn
\eea
Putting everything together, we therefore have
\bea
\phi(z,t,r,\theta,\phi) =
i\int_0^\infty d\lambda d\alpha \sum_{l,m}    \frac{\sqrt{\pi\lambda}}{2}e^{\pi\alpha/2}
z^2J_\nu(\lambda z)\frac{H_{i\alpha}^{(2)}(\lambda t)}{t}  Y_{\alpha l m}(r,\theta,\phi).
\eea
Consequently the Wightman function is
\bea
G_+(x,x^\prime) =   -\sum_{l,m} \int_0^\infty  d\lambda d\alpha   \frac{\pi\lambda}{4}
\frac{(zz^\prime)^2}{tt^\prime} J_\nu(\lambda z)J_\nu(\lambda z^\prime)  H_{i\alpha}^{(2)}(\lambda t)
H_{i\alpha}^{(1)}(\lambda t^\prime)
Y_{\alpha l m}(r,\theta,\phi)Y_{\alpha lm}(r^\prime,\theta^\prime,\phi^\prime).\nn
\eea
Using the completeness of the spherical harmonics
\bea
&& \sum_{lm} Y_{\alpha l m}(\chi, \Omega) Y_{\alpha lm}(\chi^\prime, \Omega^\prime) =
\frac{\alpha~{\rm sin} \alpha\gamma}{2\pi^2 {\rm sinh}\gamma},
\label{completeness}
\eea
we can write the Wightman function as
\bea
G_+(x,x^\prime) = \frac{\pi(zz^\prime)^2}{4tt^\prime}\int  \lambda d\lambda d\alpha
J_\nu(\lambda z)J_\nu(\lambda z^\prime)  H_{i\alpha}^{(2)}(\lambda t) H_{i\alpha}^{(1)}(\lambda t^\prime)
\frac{\alpha \sin \alpha\gamma}{2\pi^2\sinh\gamma}\nn,
\eea
where
\begin{eqnarray}
&&{\rm cosh}\gamma = {\rm cosh}\chi ~{\rm cosh}\chi^\prime - {\rm sinh}\chi ~{\rm
sinh}\chi^\prime {\rm cos}\omega,\nn\\
&&{\rm cos}\omega =
{\rm cos}\theta ~{\rm cos}\theta^\prime + {\rm sin}\theta~{\rm sin}\theta^\prime~{\rm cos}(\phi - \phi^\prime).
\label{xidef}
\end{eqnarray}
Further, doing the $\alpha$ integral, we get
\bea
G_+(x,x^\prime) =
-\frac{i(zz^\prime)^2}{4\pi^2} \int_0^\infty  \lambda^2 d\lambda    J_\nu(\lambda z)J_\nu(\lambda z^\prime)
\frac{K_1\Big( \lambda\sqrt{ -t^2 - t^{\prime 2} +
2 tt^\prime~{\rm cosh}\gamma}\Big)}{\sqrt{ t^2 + t^{\prime 2} - 2 tt^\prime~{\rm cosh}\gamma}}.\nn
\eea
Finally, after completing the $\lambda$ integral we reach at (\ref{W.M.F-5}).

\bigskip

\bigskip

\noindent\underbar{Deriving (\ref{lasteq}) and (\ref{twistedsum})}
 
\bigskip

We start with the formula \cite{gradshteyn}
\begin{equation}
\frac{1 - p^2}{1 - 2 p~{\rm cos}(\Delta \theta - \frac{2\pi k}{q}) + p^2}
= 1 + 2 \sum_{m = 1}^\infty p^m ~{\rm cos}[m(\Delta \theta - \frac{2\pi k}{q})].\nn
\end{equation}
From the above, it follows that
\begin{equation}
\sum_{k = 0}^{q-1} \frac{1 - p^2}{1 - 2 p~{\rm cos}(\Delta \theta - \frac{2\pi k}{q}) + p^2}
= q  + 2 \sum_{k= 0}^{q-1}\sum_{m=1}^\infty p^m  ~{\rm cos}[m(\Delta \theta - \frac{2\pi k}{q})].\nn
\end{equation}
Now since
\begin{equation}
\sum_{m = -\infty, \ne 0}^\infty p^{|m|} e^{im (\Delta \theta - 2\pi k/q)}
= 2 \sum_{m =1}^\infty p^m  ~{\rm cos}[m(\Delta \theta - \frac{2\pi k}{q})],\nn
\end{equation}
we get
\begin{equation}
\sum_{k = 0}^{q-1} \frac{1 - p^2}{1 - 2 p~{\rm cos}(\Delta \theta - \frac{2\pi k}{q}) + p^2}
= q + \sum_{k = 0}^{q-1} \sum_{m = -\infty, \ne 0}^\infty p^{|m|} e^{im \Delta\theta}
e^{-\frac{2i\pi m k}{q}}.\nn
\end{equation}
Further, using
\begin{equation}
\sum_{k=0}^{q-1} e^{-\frac{2i\pi m k}{q}} = q \sum_{n = -\infty}^\infty \delta_{-m, nq},\nn
\end{equation}
we arrive at
\begin{eqnarray}
\sum_{k = 0}^{q-1} \frac{1 - p^2}{1 - 2 p~{\rm cos}(\Delta \theta - \frac{2\pi k}{q}) + p^2}
&=& q +q \sum_{n = -\infty}^\infty ~~\sum_{m = -\infty, \ne 0}^\infty p^{|m|} e^{im \Delta\theta}
\delta_{-m, nq}\nonumber\\
&=& q +q \sum_{n = -\infty, \ne 0}^\infty p^{q|n|} e^{-inq\Delta\theta}\nonumber\\
&=& q  + 2 q \sum_{n = 1}^\infty p^{qn}~{\rm cos}(qn \Delta\theta)\nonumber\\
&=& q  + q\Big[ \frac{1 - p^{2q}}{1 - 2 p^q~{\rm cos}(q\Delta\theta) + p^{2q}} -1\Big]\nonumber\\
&=&  \frac{q(1 - p^{2q})}{1 - 2 p^q~{\rm cos}(q\Delta\theta) + p^{2q}}.\nn
\label{sumsum}
\end{eqnarray}
Therefore,
\begin{equation}
\sum_{k = 0}^{q-1} \frac{1}{\frac{1+p^2}{2p} - ~{\rm cos}(\Delta \theta - \frac{2\pi k}{q})}
 =  \frac{2qp(1 - p^{2q})}{(1-p^2)(1 - 2 p^q~{\rm cos}(q\Delta\theta) + p^{2q})}.\nn
\end{equation}
Taking $\frac{1+p^2}{2p} = \gamma$, we get (\ref{lasteq}).

Similarly, we can sum the series for the twisted scalar to get  (\ref{twistedsum}). 
Since $\beta=0,1$ cases reduce to untwisted scalar case, we will restrict to $0 < \beta <1$. Again, we start with 
\begin{equation}
\sum_{k = 0}^{q-1} \frac{e^{-2\pi i \beta k}(1 - p^2)}{1 - 2 p~{\rm cos}(\Delta \theta - \frac{2\pi k}{q}) + p^2}
= \sum_{k = 0}^{q-1} e^{-2\pi i \beta k} + 2 \sum_{k= 0}^{q-1}\sum_{m=1}^\infty e^{-2\pi i \beta k}p^m  ~{\rm cos}[m(\Delta \theta - \frac{2\pi k}{q})].\nn
\end{equation}

First sum on the RHS vanishes for the range of $\beta$ that we are considering while the second term can be evaluated using the same methods as before to get 
\begin{equation}
\sum_{k = 0}^{q-1} \frac{e^{-2\pi i \beta k}(1 - p^2)}{1 - 2 p~{\rm cos}(\Delta \theta - \frac{2\pi k}{q}) + p^2}
=  q \sum_{n = -\infty, \ne -\beta}^\infty p^{q|-n-\beta|} e^{-i(nq+ q\beta)\Delta\theta}\nonumber\\ .\nn
\end{equation}

Since $n$ takes integer values while $\beta$ is not an integer, there is no restriction on the sum. Expanding the sum, we get
\be
\sum_{k = 0}^{q-1} \frac{e^{-2\pi i \beta k}(1 - p^2)}{1 - 2 p~{\rm cos}(\Delta \theta - \frac{2\pi k}{q}) + p^2}
=  q p^{q\beta} e^{-i( q\beta)\Delta\theta} + q
 \sum_{n = -\infty }^1 p^{q|-n-\beta|} e^{-i(nq+ q\beta)\Delta\theta} + q \sum^{ \infty }_{n=1} p^{q|-n-\beta|} e^{-i(nq+ q\beta)\Delta\theta} 
 \nonumber\\ 
\ee

Relabelling $n\rightarrow -n$ in the second term, we get 

\be
\sum_{k = 0}^{q-1} \frac{e^{-2\pi i \beta k}(1 - p^2)}{1 - 2 p~{\rm cos}(\Delta \theta - \frac{2\pi k}{q}) + p^2}
=  q p^{q\beta} e^{-i q\beta\Delta\theta} + q e^{-i q\beta\Delta\theta}\left( p^{-q\beta }
  \sum^{ \infty }_{n=1} p^{q n} e^{inq\Delta\theta} + p^{q\beta }
  \sum^{ \infty }_{n=1} p^{q n} e^{-inq\Delta\theta}\right)
   \nonumber\\ 
\ee
Now we can do the final sums using the formula 
\be
\sum_{n=0}^{\infty} s^n e^{in x} = \frac{1-s e^{-ix}}{1-2s \cos x + s^2} \nonumber
\ee

\begin{eqnarray}
\sum_{k = 0}^{q-1} \frac{e^{-2\pi i \beta k}(1 - p^2)}{1 - 2 p~{\rm cos}(\Delta \theta - \frac{2\pi k}{q}) + p^2}
=  q p^{q\beta} e^{-iq\beta\Delta\theta} + qp^{-q\beta } e^{-i q\beta\Delta\theta}\left( \frac{1- p^q e^{-i q\Delta\theta}}{1-2p^q \cos( q\Delta \theta) + p^{2q}} -1\right) \nonumber \\
+  qp^{q\beta } e^{-i q\beta\Delta\theta}\left( \frac{1- p^q e^{i q\Delta\theta}}
{1-2p^q \cos (q\Delta \theta) + p^{2q}} -1\right) \nonumber
\end{eqnarray}

Simplifying we finally get
\begin{equation}
\sum_{k = 0}^{q-1} \frac{e^{-2\pi i \beta k}}{\frac{1+p^2}{2p} - ~{\rm cos}(\Delta \theta - \frac{2\pi k}{q})}
 =  \frac{2qp e^{-iq\beta\Delta\theta} (p^{q\beta} - p^{2q} -2p^q e^{iq\Delta\theta}\sinh( q\beta \ln p))}{(1-p^2)(1 - 2 p^q~{\rm cos}(q\Delta\theta) + p^{2q})}.\nn
\end{equation}
Taking $\frac{1+p^2}{2p} = \gamma$, we get (\ref{twistedsum}).
\bigskip

\bigskip

\noindent\underbar{Identities used in the main text}

\bigskip

Here we collect a number of identities used in the main text.

First, to get the normailzation of $C_{\lambda\alpha n}$, we have used the following properties of
$P_{i\alpha-1/2}^{-n} (x)$ \cite{zhurina}:
\begin{eqnarray}
P_{iz - 1/2}^{m}(u) = \frac{\Gamma[-iz +m +1/2]}{\Gamma[-iz - m + 1/2]} P_{iz - 1/2}^{-m}(u)
= \frac{\Gamma[iz +m +1/2]}{\Gamma[iz - m + 1/2]} P_{iz - 1/2}^{-m}(u)\nonumber
\end{eqnarray}
and
\begin{eqnarray}
\int_1^\infty du P_{i z' - 1/2}^{-m} (u) P_{i z - 1/2}^{-m} (u) = \frac{\pi}{z ~{\rm sinh}(\pi z)}
\frac{\delta(z - z')}{|\Gamma(m + 1/2 + i z)|^2}.
\label{norma2}
\end{eqnarray}
}

To get to (\ref{midexp}), we have used
\begin{eqnarray}
H_{i\alpha}^{(2)*}(x) = H_{-i\alpha}^{(1)},
~H_{i\alpha}^{(2)}(\lambda t) H_{-i\alpha}^{(1)}(\lambda t') = \frac{4}{\pi^2} e^{-\pi\alpha}
K_{i\alpha}(i\lambda t) K_{-i\alpha}(-i\lambda t'),\nonumber
\end{eqnarray}
and
\begin{eqnarray}
&&2\sum_{n=1}^\infty (-)^n ~{\rm cos}n(\theta - \theta')
P_{i\alpha - 1/2}^{-n} ({\rm{cosh}}r)P_{i\alpha -
1/2}^{n}
({\rm{cosh}}r') \nonumber\\
&&~~~~~~~~~+ P_{i\alpha - 1/2} ({\rm{cosh}}r)P_{i\alpha - 1/2} ({\rm{cosh}}r')
= P_{i\alpha - 1/2}({\rm{cosh}}\chi).
\label{appen1}
\end{eqnarray}

To reach to equation (\ref{bulkwightman}), we used \cite{gradshteyn}
\begin{eqnarray}
&&\int_0^\infty d\alpha~\alpha ~{\rm sinh}(\pi \alpha )\frac{\pi}{{\rm cosh}\pi \alpha}K_{i\alpha}(a)
K_{i\alpha}(b) P_{i\alpha -\frac{1}{2}} ({\rm cosh}\chi) \nonumber\\
&&~~~~~= \sqrt{\frac{\pi}{2}} \Big(\frac{ab}{\sqrt{a^2 + b^2 + 2 a b ~{\rm cosh}\chi}}\Big)^{\frac{1}{2}}
K_{\frac{1}{2}}
(\sqrt{a^2 + b^2 + 2 a b ~{\rm cosh}\chi}).
\label{appen2}
\end{eqnarray}
This identity is valid when $|arg(a)| < \pi/2$ and so we replace $t$ with $t e^{-i\epsilon}$, 
with limit $\epsilon \rightarrow 0$.

Equation (\ref{cdelta}) uses the identity
\begin{eqnarray}
\int_0^\infty \frac{{\rm cosh}2yt}{{\rm cosh}^{2x} bt} dt = \frac{4^{x-1}}{b} B(x+y/b, x-y/b),
\label{appen3}
\end{eqnarray}
Here $B(x,y)$ represents the beta function.

We used relation \cite{Bailey} 
\begin{eqnarray}
&&|\Gamma(i\alpha + n q + 1/2)|^2  P_{i\alpha -1/2}^{-qn}({\rm cosh}r) P_{i\alpha -1/2}^{-qn}({\rm
cosh}r') \nonumber\\
&&~~~~~= \sqrt{\frac{2}{\pi}} \int_0^\infty dx x^{-1/2} I_{|n|q} (x ~{\rm sinh}r~{\rm sinh}r') K_{i\alpha}(x)
e^{-x ~{\rm cosh}r~{\rm cosh}r'},\nonumber
\end{eqnarray}
where $I_\nu(x)$ is the modified Bessel function and \cite{Oberhettinger}
\begin{eqnarray}
\int_0^\infty x~{\rm sinh}(\pi x) K_{ix}(a) K_{ix}(b) K_{ix}(y) dx = \frac{\pi^2}{4} e^{-\frac{y}{2}
\Big(\frac{a}{b} + \frac{b}{a} + \frac{ab}{y^2}\Big)},
\label{appen4}
\end{eqnarray}
to get (\ref{bbbulk}). 

Using \cite{gradshteyn}
\begin{eqnarray}
\int_0^\infty e^{-\alpha x} J_\nu(2\beta \sqrt{x}) J_\nu(2 \gamma\sqrt{x}) dx = \frac{1}{\alpha} 
I_\nu\Big(\frac{2\beta \gamma}{\alpha}\Big) e^{-\Big(\frac{\beta^2 + \gamma^2}{\alpha}\Big)},\label{lambdaintegral}
\end{eqnarray}

we can perform the $\lambda$ integral in (\ref{bbbulk})to get (\ref{detail1}).  

Using the known result \cite{gradshteyn}
\begin{eqnarray}
\int_0^\infty e^{-x u} I_\nu(x) x^{\mu -1} dx = \sqrt{\frac{2}{\pi} } 
\frac{e^{-i(\mu - 1/2)\pi}}{\sqrt{u^2 -1}} Q_{\nu -1/2}^{\mu -1/2}(u),\label{xintegral}
\end{eqnarray}

we perform the integral over $s$ to go from (\ref{detail1} )to (\ref{renW}).

\end{document}